 \documentclass[twocolumn,prb]{revtex4}

\usepackage{graphicx}    % Include figure files
\usepackage{dcolumn}     % Align table columns on decimal point
\usepackage{bm}          % bold math 

\begin{document}

\title{Structural, electronic and magnetic properties of vacancies
in single-walled carbon nanotubes}

\author{W. Orellana and P. Fuentealba}
\affiliation{Departamento de F\'{\i}sica, Facultad de Ciencias, 
Universidad de Chile, Casilla 653, Santiago, Chile.}

%date\today

\begin{abstract} 
The stability and properties of the monovacancy and the divacancy in 
single-walled carbon nanotubes (CNTs) are addressed by spin-density functional 
calculations. We study these defects in four nanotubes, the armchair (6,6) 
and (8,8) and the zigzag (10,0) and (14,0), which have diameters 
of about 8 and 11~\AA, respectively. We also study different defect 
concentrations along the tube axis by increasing the supercell in this
direction in order to have one defect every 13 and 26~\AA\ of CNT length. 
Our results show that in the equilibrium geometry CNTs with a monovacancy
exhibit ferromagnetism with magnetic moments ranging from 0.3 to 
0.8~$\mu_{\rm B}$. On the other hand, CNTs with a divacancy do not exhibit
magnetism due to the full reconstruction around the defect where all C atoms
are three coordinated.
We observe that the presence of a monovacancy does not change drastically 
the CNT electronic properties, preserving their corresponding metallic or 
semiconducting character. However, for the divacancy both armchair and
zigzag CNT become semiconductors exhibiting a energy gap of about 0.15~eV.
\end{abstract} 

%\keywords{Carbon Nanotubes, Monovacancy, Divacancy, Magnetic and 
% Electronic Properties, Density Functional Calculations}

\maketitle
\section{Introduction}
The study of radiation-induced defects in carbon nanostructures have become 
considerably important since the recent experimental results of 
room-temperature magnetic ordering in proton-irradiated graphite, which 
exclude metal impurities\cite{esquinazi}, as well as the observation of 
ferromagnetism in pressure-induced polymerized fullerenes \cite{makarova,wood}. 
Magnetic ordering in defective graphite and fullerenes have been also 
predicted theoretically\cite{lehtinen,kim}, suggesting the possibility to 
realize nanoscale magnets with potential applications in spintronics 
and biocompatible magnetic materials. 
On the other hand, magnetism in single-walled carbon nanotubes 
(CNTs) appears to be more complex than graphite and fullerenes owing the 
curvature, chirality and different electronic characteristics of the tubes. 
Single-walled CNTs, which are typically 10~\AA\ in diameter, can be metallic 
or semiconducting depending on the way that a graphene sheet is rolled up, 
which is characterized by the chiral indices ($n,m$). Metallic tubes occur 
if $n-m$ is divisible by 3, otherwise the tubes are semiconducting
\cite{dressel}. Tubes with indices ($n,0$) and ($n,n$) are termed 
$zigzag$ and $armchair$, respectively, which is related to the arrangement 
of carbon atoms around the tube.

Experimental knowledge concerning defect-induced magnetism in CNTs is still
lacking. The difficult to measure magnetic ordering in nanotube samples 
resides in the presence of metal catalysts which are necessary to produce 
single-walled tubes. Theoretical investigations on open ended zigzag
nanotubes are found to exhibit energetically favorable ferromagnetic spin 
configurations which are associated to unsaturated dangling bonds at zigzag 
edges\cite{kim}. Single vacancies in small-diameters CNTs, ranging 
from 4 to 8~\AA, have been recently addressed by first-principles calculations
\cite{ma}. Although relative high concentrations of vacancy or small supercell
sizes containing a vacancy have been considered in this work, it shows that 
only metallic nanotubes with a single vacancy in their ground state equilibrium
geometries would exhibit ferromagnetism.

%In this work we focus on the stability, electronic and magnetic properties of 
%the monovacancy and the divacancy in armchair and zigzag single-walled CNTs 
%with diameters of about 8 and 11~\AA\ as well as two defect concentrations
%$of one defect every 13 and 26~\AA\ of CNT length. We find that both armchair 
%and zigzag CNTs containing a monovacancy may give rise to magnetic moments 
%among 0.3 to 0.8~$\mu_{\rm B}$, depending on the diameters and defect 
%concentrations. On the other hand, CNTs with a divacancy would not 
%exhibit magnetism. Our results suggest interesting applications for defective
%CNTs in molecular magnets and ferromagnetic arrays.

\section{Theoretical approach}
The calculations were performed in the framework of the spin-polarized 
density functional theory \cite{kohn}, using a basis set of 
strictly-localized numerical pseudoatomic orbitals as implemented in the 
SIESTA code \cite{siesta}. 
The exchange-correlation energy is calculated within the generalized gradient
approximation \cite{perdew}.
Standard norm-conserving pseudopotentials \cite{troullier} in their separable 
form \cite{klein} are used to describe the electron-ion interaction.
We use a double-$\zeta$ singly-polarized basis set. We study the vacancies
in both armchair and zigzag CNTs with different diameters. The armchair CNTs 
have chiral indices of (6,6) and (8,8), whereas the zigzag ones have indices 
of (10,0) and (14,0). For the (6,6) and (10,0) CNTs which have diameters of 
about 8~\AA, we use unit cells containing 120 atoms. 
For the (8,8) and (14,0) CNTs with diameters of about 11~\AA, the 
supercells contain 160 and 168 atoms, respectively. For both 8 and 11~\AA\
diameter CNTs the supercell has a length of $L \approx 13$~\AA. 
This means that we are calculating one defect every 13~\AA\ of CNT length.
For the Brillouin zone 
sampling we use six {\it k} points along the CNT axis \cite{monk}. To ensure
negligible interaction between tubes we imposed a vacuum region of 12~\AA.
With the above supercells, the distance between vacancy images along the tube
axis is about 13~\AA. This means that we are simulating a CNT with a linear 
concentration of defects with one defect every 13~\AA\ of CNT length.
A longer supercell with length of $2L \approx 26$~\AA\ are also consider to 
study the effect a lower defect density in the CNT electronic structure. 
To do that we use (6,6) and (10,0) nanotubes in supercell containing up to
240 atoms and the $\Gamma$ point for the BZ sampling.
The positions of all atoms in the nanotubes were relaxed using the conjugated
gradient algorithm until the force components become smaller than 
0.05~eV/$\text{\AA}$.

\section{The monovacancy in carbon nanotubes}

Figures~1(a) and 1(b) show respectively the equilibrium geometries 
of the (6,6) and (10,0) CNTs containing the monovacancy (hereafter referred 
to as (6,6)+1V and (10,0)+1V). 
%%%%%%%%%%%%%%%%%%%%%%%%%%%%%%%%%%%%%%%%%%%%%%%%%%%%%%%%%%%%%%%%%%
\begin{figure}
\includegraphics[width=7.5cm]{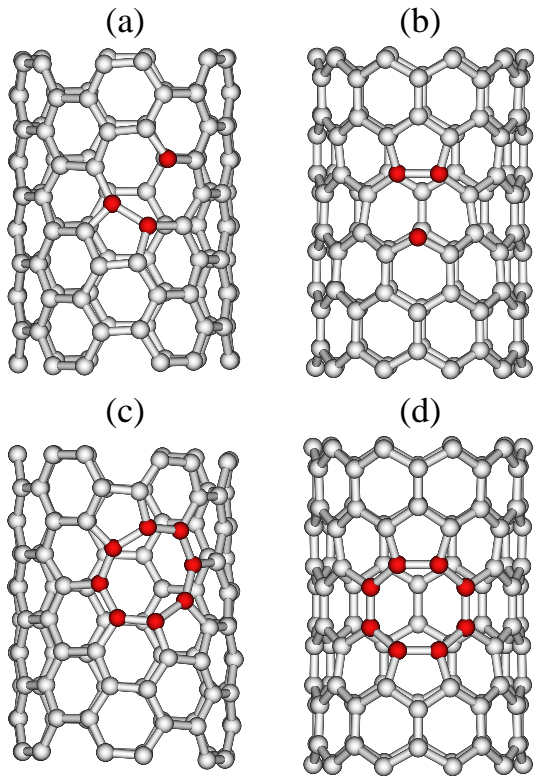}
\caption{\label{f1} Equilibrium geometry of the monovacancy and the
divacancy in (6,6) and (10,0) CNTs.
(a) (6,6)+1V,
(b) (10,0)+1V,
(c) (6,6)+2V,
(d) (10,0)+2V.
Painted balls display the C atoms around the defect.}
\end{figure}
%%%%%%%%%%%%%%%%%%%%%%%%%%%%%%%%%%%%%%%%%%%%%%%%%%%%%%%%%%%%%%%%%%
We find that the neighboring C atoms partially
reconstruct around the defect, forming a pentagon bonding structure, leaving a 
C atom twofold coordinated, which is projected slightly off to the nanotube surface. 
We also find metastable configurations for (6,6)+1V and (10,0)+1V, 10.5 and 
12.1~meV/atom higher in energy that their respective ground state (GS) geometries. 
In these configurations the pentagon structure plus the twofold coordinated C
atom are also found but rotated clockwise about $60^{\circ}$ with respect to the 
equilibrium geometries.
A third metastable structure is found for (10,0)+1V, which is 17.2~meV/atom higher 
in energy than the GS geometry. Here, the C atoms neighboring to the vacancy do 
not form bond, remaining twofold coordinated. This geometry is known as $3db$ 
(three dangling bonds) structure.

Similar results for the GS and metastable geometries are found for the 
vacancy in the 11~\AA\ diameter CNTs (8,8) and (14,0). The metastable 
geometries of (8,8)+1V and (14,0)+1V are 7.3 and 6.7~meV/atom higher in energy 
that the respective GS geometries. The smaller energy difference with respect 
to the 8~\AA\ diameter CNTs is attributed to a decrease in the strain energy 
due to the lower curvature of the (14,0) and (8,8) tubes. Thus, extrapolating 
to higher-diameter tubes, we can infer that the metastable geometries would 
tend to disappear. It is interesting to note that the $3db$ metastable
structure of (10,0)+1V is not found in (14,0)+1V, suggesting that the $3db$ 
geometry exits due to the higher strain in the (10,0) tube, hindering further 
relaxations.

We calculate the formation energy of a vacancy in our CNTs using,
\begin{equation}
 E_{form}(V) = E_{tot}({\rm CNT+V})-E_{tot}({\rm CNT})+\mu_{C} 
\end{equation}
where $E_{tot}$(CNT+V) and $E_{tot}$(CNT) are the total energy of the tube 
containing the vacancy and the perfect tube, respectively. $\mu_{C}$ is the 
carbon chemical potential which is calculated as the total energy per atom in
the perfect tube. Our results are listed in Table~I where we also include 
vacancy calculations performed in (6,6) and (10,0) CNTs considering a supercell
with twice the length of the one previously used. 
%%%%%%%%%%%%%%%%%%%%%%%%%%%%%%%%%%%%%%%%%%%%%%%%%%%%%%%%%%%%%%%%%%
\begin{table}[b]
\caption{\label{t1} Formation energy ($E_{form}$) and magnetic moment 
($m$) for the monovacancy and the divacancy in armchair and zigzag CNTs, 
considering different supercell length ($L$) or vacancy concentrations.
$d_{\rm C-C}$ is the equilibrium distance between C atoms that 
approach each other to form the pentagon bonding structure. In parenthesis 
are shown our results for the metastable geometries.}
\begin{ruledtabular}
\begin{tabular}{lcllll}
System&$L$ (\AA)&$m$ ($\mu_B$)&$E_{form}$ (eV)&$d_{\rm C-C}$ (\AA)\\
\hline \\
(6,6)+1V  & 12.46 & 0.82~(1.1) & 5.75~(7.00) & 1.56~(1.93) \\
(6,6)+1V  & 24.92 & 0.63       & 5.85        & 1.57        \\
(10,0)+1V & 12.96 & 0.32~(1.0) & 5.67~(7.11) & 1.54~(1.72) \\
(10,0)+1V & 25.92 & 0.49       & 5.65        & 1.53        \\
(8,8)+1V  & 12.46 & 0.77~(1.1) & 4.65~(5.81) & 1.58~(1.98) \\
(14,0)+1V & 12.96 & 0.53~(1.1) & 5.20~(6.32) & 1.58~(1.74) \\
(6,6)+2V  & 12.46 & 0.0~(0.0) & 4.24~(7.84) & 1.53~(1.78) \\
(6,6)+2V  & 24.92 & 0.0       & 4.17        & 1.52        \\
(10,0)+2V & 12.96 & 0.0~(0.0) & 3.90~(6.65) & 1.50~(1.65) \\
(10,0)+2V & 25.92 & 0.0       & 4.06        & 1.49        \\
(8,8)+2V  & 12.46 & 0.0       & 2.09        & 1.53        \\
(14,0)+2V & 12.96 & 0.0       & 2.33        & 1.50        \\
\end{tabular}
\end{ruledtabular}
\end{table}
%%%%%%%%%%%%%%%%%%%%%%%%%%%%%%%%%%%%%%%%%%%%%%%%%%%%%%%%%%%%%%%%%%
We find that the vacancy in the 8~\AA\ diameter CNTs have similar formation 
energies, ranging from 5.65 to 5.85~eV, including the longer supercell. 
The metastable geometries also show the same tendency, with formation 
energies of about 7~eV.
The above suggests a minor influence between images vacancies for supercell 
with lengths up to $L\approx$13~\AA. 
For the vacancy in CNTs of 11~\AA\ of diameter we find lower formation energies 
than those of the 8~\AA\ ones, where the defective tube (8,8)+1V [(14,0)+1V] has a 
formation energy of about 1~eV [0.5~eV] lower than that found for (6,6)+1V 
[(10,0)+1V]. The decrease in the formation energy is attributed to the decrease 
in the strain energy due to the small curvature of the larger tubes.

Table~I shows our results for the magnetic moment of the defective CNTs. We
find that all nanotubes under consideration exhibit ferromagnetic ordering.
The lower-energy structures of (6,6)+1V and (10,0)+1V have magnetic moments of 
0.82 and 0.32~$\mu_{\rm B}$, respectively, whereas for the metastable 
configuration the magnetic moment are closer to 1~$\mu_{\rm B}$. Clearly,
the magnetism is due to the under-coordinated C atom which has a localized
unpaired spin, as shown in Fig.~2. 
%%%%%%%%%%%%%%%%%%%%%%%%%%%%%%%%%%%%%%%%%%%%%%%%%%%%%%%%%%%%%%%%%%
\begin{figure}
\includegraphics[width=7.0cm]{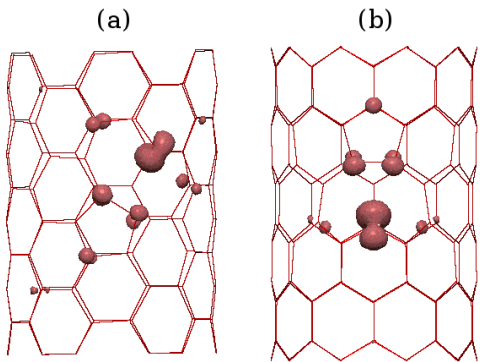}
\caption{\label{f2} Spin-density isosurfaces for the monovacancy
in 8~\AA\ diameter CNTs. (a) (6,6)+1V and (b) (10,0)+1V. 
The isosurfaces correspond to a charge of 0.015 $e$/\AA$^3$.} 
\end{figure}
%%%%%%%%%%%%%%%%%%%%%%%%%%%%%%%%%%%%%%%%%%%%%%%%%%%%%%%%%%%%%%%%%%
However, as a C atom with a dangling bond has a magnetic
moment of 1~$\mu_{\rm B}$, the lower magnetic moments found in 
(6,6)+1V and (10,0)+1V are due to a redistribution of charge around the
defect. This can be checked by looking at the equilibrium distance between the 
C atoms that form the pentagon structure, shown in Table~I. As longer 
this distance is, larger is the magnetic moment. This means that once a vacancy 
is created, two of the three dangling bonds containing a spin rehybridize 
forming the pentagon structure. The weakness of this bond depend on the curvature
strain of the tube. An extreme case occurs in the $3db$ configuration 
of (10,0)+1V. Here no bond between the under-coordinated C atoms is formed, 
resulting in three dangling bonds with magnetic moment of 1.4~$\mu_{\rm B}$.
Considering lower concentration of vacancies, that is doubling the supercell
length, we find that magnetic moments on (6,6)+1V and (10,0)+1V are 0.63 and 
0.49~$\mu_{\rm B}$, respectively. Thus, the magnetic moment tend 
to stabilize approximately around the average value 0.6~$\mu_{\rm B}$ for more
diluted monovacancy concentrations. The same behavior is observed for the tubes with 
higher diameters where the magnetic moment on (8,8)+1V and (14,0)+1V are 0.77 and 
0.53~$\mu_{\rm B}$, respectively. This stabilization appears to be related to 
a decrease in the strain energy on the tube surface. Thus, we may infer that the 
magnetization due to a single vacancy would be present in a wide variety of CNTs. 

Figures~3(b) and 3(c) show the spin-resolved density of states for (6,6)+1V 
CNT calculated with both supercells length (12.46 and 24.96~\AA\ respectively),
which are compared with density of states of perfect (6,6) CNT [Fig.~3(a)]. 
%%%%%%%%%%%%%%%%%%%%%%%%%%%%%%%%%%%%%%%%%%%%%%%%%%%%%%%%%%%%%%%%%%
\begin{figure}
\includegraphics[width=6.25cm]{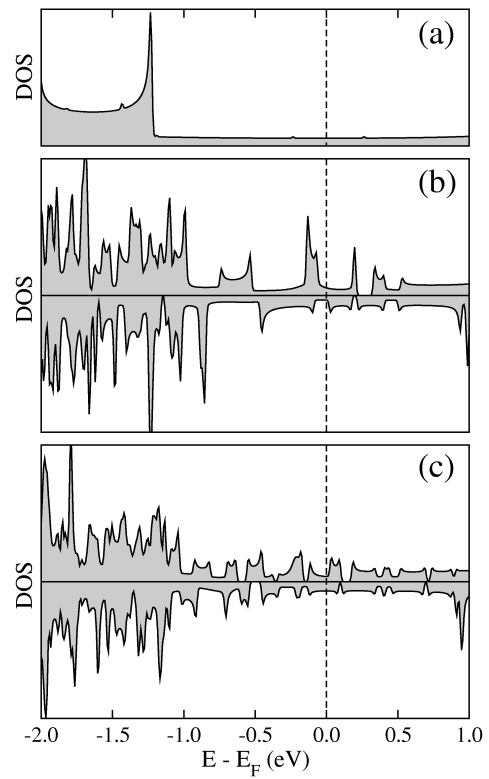}
\caption{\label{f3} Spin-resolved density of states (DOS) for the (6,6) 
CNT with a monovacancy.
(a) The perfect (6,6) CNT
(b) (6,6)+1V calculated with a supercell of 12.46~\AA\ length.
(c) (6,6)+1V calculated with a supercell of 24.96~\AA\ length.
In (a) and (b), upper and lower panels correspond to majority and
minority spin configurations, respectively.
Dashed lines indicate the position of the Fermi level (E$_{\rm F}$).}
\end{figure}
%%%%%%%%%%%%%%%%%%%%%%%%%%%%%%%%%%%%%%%%%%%%%%%%2%%%%%%%%%%%%%%%%%%
We observe that the (6,6) armchair CNT preserves its metallic character for
different vacancy concentrations. We also note that the difference between
majority (upper panel) and minority (lower panel) spin configurations in
Fig.~3(b) and 3(c) clearly demonstrate the magnetism in the (6,6)+1V system. 
Similar results are found for (8,8)+1V CNT which suggests that these properties 
would be also found in larger diameter armchair CNTs. Figures~4(b) and 4(c) 
show the spin-resolved density of states for (10,0)+1V CNT calculated with 
supercells of 12.46 and 24.96~\AA\ length, and the density of states of
perfect (10.0) CNT [Fig.~4(a)]. As we can see the (10,0) semiconducting zigzag
CNTs maintain this property when a C atoms is removed. For higher vacancy
concentration the gap energy is about 0.1~eV mainly due to a state located
just above the Fermi energy. For lower vacancy concentration this state 
is less dispersive increasing the gap energy up to 0.2~eV. The presence of
this states give to the (10,0)+1V CNT an acceptor character.
%%%%%%%%%%%%%%%%%%%%%%%%%%%%%%%%%%%%%%%%%%%%%%%%2%%%%%%%%%%%%%%%%%%
\begin{figure}
\includegraphics[width=6.25cm]{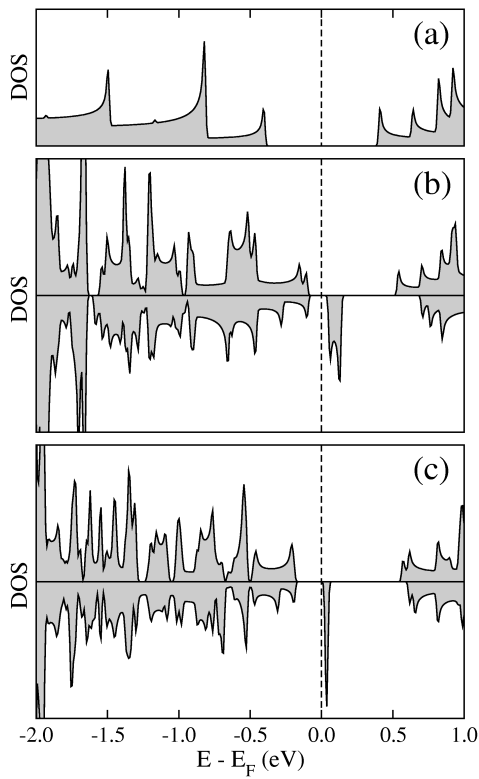}
\caption{\label{f4} Spin-resolved density of states for the (10,0)
CNT with a monovacancy.
(a) The perfect (10,0) CNT
(b) (10,0)+1V calculated with a supercell of 12.96~\AA\ length.
(c) (10,0)+1V calculated with a supercell of 25.92~\AA\ length.
Dashed lines indicate the position of the Fermi Level (E$_{\rm F}$).}
\end{figure}
%%%%%%%%%%%%%%%%%%%%%%%%%%%%%%%%%%%%%%%%%%%%%%%%%%%%%%%%%%%%%%%%%%

\section{The divacancy in carbon nanotubes}

Figures~1(c) and 1(d) display the GS geometries for the divacancy in 
(6,6) and (10,0) CNTs (hereafter referred to as (6,6)+2V and (10,0)+2V).
We also find metastable geometries for (6,6)+2V and (10,0)+2V, which
are 10.5 and 23.3~meV/atom higher in energy than the respective GS 
geometries. We observe that once a divacancy is created, the 
undercoordinated C atoms spontaneously reconstruct  around the defect 
forming an octagon with two adjacent pentagon in opposite directions. 
For (6,6)+2V the pentagons point along a line of about 30$^{\circ}$ 
with respect to the tube axis, whereas in the metastable structure the 
pentagons are aligned normal to the tube axis. 
For (10,0)+2V the pentagons align along the tube axis 
for the GS geometry and along a line of about 60$^{\circ}$ with 
respect the tube axis. Similar reconstructed geometries for (8,8)+2V 
and (14,0)+2V CNTs where found, suggesting that these reconstructions 
should be found in larger diameters CNTs. For the 8~\AA\ (6,6) and 
(10,0) CNTs the divacancy formation energies are found to be 4.2 and 
4.0~eV, respectively. Using longer supercell we note a small change in 
the formation energies suggesting a negligible interaction between the 
defect and their images in neighboring supercell. For the 11~\AA\ (8,8) 
and (14,0) CNTs the formation energies decrease to 2.1 and 2.3~eV, 
respectively, about half the value found in the 8~\AA\ CNTs. This suggests
that the curvature strain becomes important for the divacancy formation in 
small diameter CNTs. In Table~I we also show the bond length between 
the C atoms that form the pentagon structure. The smaller bond
length, of about 1.5~\AA, are found for the divacancy in the zigzag (10,0) 
and (14,0) CNTs. 
As a consequence of the full reconstruction of the divacancy all C atoms
around the octagon are threefold coordinated. Thus, no unpaired $\sigma$
electron is found, implying that carbon nanotubes with this defect do not 
exhibit magnetism. Therefore, we can infer that the origin of the magnetism 
in nanotubes and graphite are mainly due to unpaired $\sigma$ electrons which 
are localized. On the other hand, the delocalized $\pi$ electrons would have a 
negligible contribution because once an unpaired $\sigma$ electron reconstructs 
the magnetization disappear as shown our divacancy results.

Figures~5(b) and 5(c) shows the density of states for the (6,6)+2V CNT 
calculated with supercells of 12.46 and 24.96~\AA\ length, respectively, which 
are compared with the density of states of the perfect (6,6) CNT [Fig.~5(a)]. 
%%%%%%%%%%%%%%%%%%%%%%%%%%%%%%%%%%%%%%%%%%%%%%%%%%%%%%%%%%%%%%%%%%
\begin{figure}
\includegraphics[width=6.25cm]{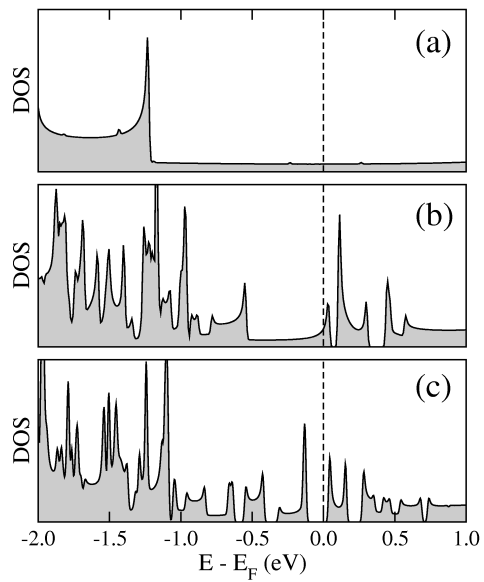}
\caption{\label{f5} Density of states for the (6,6) CNT with a divacancy. 
(a) Perfect (6,6) CNT
(b) (6,6)+2V calculated with a supercell of 12.46~\AA\ length.
(c) (6,6)+2V calculated with a supercell of 24.96~\AA\ length.
Dashed lines indicate the position of the Fermi Level (E$_{\rm F}$).}
\end{figure}
%%%%%%%%%%%%%%%%%%%%%%%%%%%%%%%%%%%%%%%%%%%%%%%%%%%%%%%%%%%%%%%%%%
We find that for the higher defect concentration the system has a metallic
character with a small energy gap just above the Fermi energy, whereas for the 
lower concentration it become a semiconductor with a energy gap of about 
0.15~eV. This behavior is originated in a dispersive CNT state which cross the 
Fermi level at about 0.8$\Gamma$X. In the more diluted defect concentration 
this state becomes flat opening a energy gap and turning the system a 
semiconductor. Similar results are found for the (10,0)+2V CNT shown in Fig.~6.
Here, the perfect (10,0) CNT is a semiconductor with an energy gap of 0.8~eV
[see Fig.~6(a)].
%%%%%%%%%%%%%%%%%%%%%%%%%%%%%%%%%%%%%%%%%%%%%%%%%%%%%%%%%%%%%%%%%%
\begin{figure}[t]
\includegraphics[width=6.25cm]{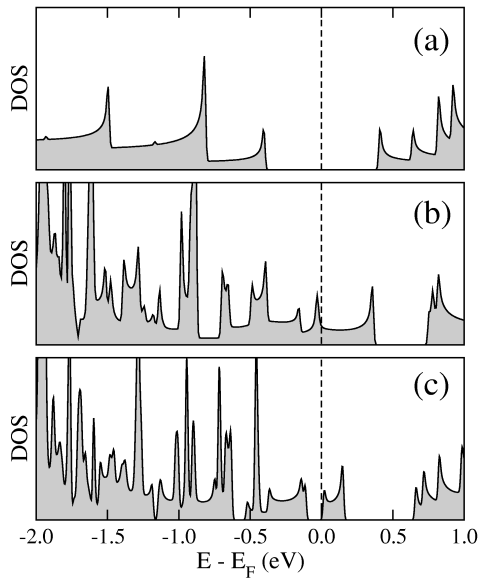}
\caption{\label{f6} Density of states for the (10,0) CNT with a divacancy. 
(a) Perfect (10,0) CNT
(a) (10,0)+2V calculated with a supercell of 12.96~\AA\ length.
(b) (10,0)+2V calculated with a supercell of 25.92~\AA\ length.
Dashed line indicate the position of the Fermi Level (E$_{\rm F}$).}
\end{figure}
%%%%%%%%%%%%%%%%%%%%%%%%%%%%%%%%%%%%%%%%%%%%%%%%%%%%%%%%%%%%%%%%%%
For the higher vacancy concentration [Fig.~6(b)] a defect state crosses the 
Fermi level at about 0.1$\Gamma$X while a CNT state rises close to the 
Fermi level, giving a metallic character to the system. However, for the more
dilute defect concentration, the (10,0)+2V CNT becomes semiconducting mainly
due to the defect state above the Fermi level which becomes flat opening
a energy gap of about 0.15~eV as shown in Fig.~6(c). It is interesting
to note that the (8,8) and (14,0) CNTs with a divacancy, calculated 
with supercells of $L\approx13$~\AA, exhibit an energy gap of about 0.15~eV. 
This suggests that for more diluted defect concentrations in CNTs with diameters
of at least 11~\AA, the gap energy should increase due to the flattening of 
the defect states close to the Fermi energy.

\section{summary and conclusions}

We have studied the monovacancy and divacancy in armchair and zigzag CNTs
with diameters of about 8 and 11~\AA\ and considering two concentrations
of defects along the tube axis. Our results show that the CNTs with a 
monovacancy exhibits a ferromagnetic ordering induced by an undercoordinated
C atom. The magnetic moment in the defect equilibrium geometries fluctuate 
among 0.3 and 0.8~$\mu_{\rm B}$, depending on the defect concentration and 
on the diameter of the CNTs. The electronic density of states along the CNTs
axis show that the monovacancy does not change the electronic character of 
the tubes. On the other hand, once a divacancy is created the neighboring 
C atoms spontaneously reconstruct around the defect forming an octagon 
with two adjacent pentagon in opposite directions. Because all the C atoms
are three coordinated in this defect, CNTs with a divacancy are not magnetic.
However, the electronic properties of CNTs change when they contain a divacancy
where both armchair and zigzag CNTs become semiconductor having a energy gap
of about 0.15 eV.
The above properties suggest the possibility to achieve nanoscale magnets 
and magnetic patterns with potential applications in spintronics and biocompatible 
magnetic materials.

\acknowledgments
We thank Millennium Nucleus of Applied Quantum Mechanics and Computational 
Chemistry for financial support, under project P02-004-F. WO also
thanks FONDECYT for partially support this work under project No. 1050197.


\begin{thebibliography}{25}

\bibitem{esquinazi}
P. Esquinazi, D. Spemann, R. H\"ohne, A. Setzer, K.-H. Han, and T. Butz,
Phys. Rev. Lett. {\bf 91} (2004) 227201.

\bibitem{makarova} 
T. Makarova, B. Sundqvist, P. Esquinazi, {\it et al.}, 
Nature {\bf 413} (2001) 718.

\bibitem{wood} 
R.A. Wood, M.H. Lewis, M.R. M\"uller, D. Eckert, {\it et al.},
J. Phys.: Condens. Matter {\bf 14} (2002) L385.

\bibitem{lehtinen} 
P.O. Lehtinen, A.S. Foster, Y. Ma, A.V. Krasheninnikov, 
and R.M. Nieminen, Phys. Rev. Lett. {\bf 93} (2004) 187202.

\bibitem{kim} 
Y.-H. Kim, J. Choi, K.J. Chang, and D. Tom\'anek, Phys. Rev. B
{\bf 68} (2002) 125420.

\bibitem{dressel} M.S. Dresselhaus, G. Dresselhaus, and Ph. Avouris,
{\it Carbon nanotubes: Synthesis, Structure, Properties and Applications}
(Springer-Verlag, New York, 2001).

\bibitem{ma} 
Y. Ma, P.O. Lehtinen, A.S. Foster, and R.M. Nieminen, 
New J. Phys. {\bf 6} (2004) 68.

\bibitem{kohn} 
W. Kohn, Rev. Mod. Phys. {\bf 71} (1999) 1253.

\bibitem{siesta}
P. Ordej\' on, E. Artacho, and J. M. Soler, Phys. Rev. B {\bf 53} 
(1996) R10441; J.M. Soler, E. Artacho, J.D. Gale, A. Garc\'{\i}a,
J. Junquera, P. Ordej\'on, D. S\'anchez-Portal, J. Phys.: Condens.
Matter {\bf 14} (2002) 2745.

\bibitem{perdew}
J.P. Perdew, K. Burke, and M. Ernzerhof, Phys. Rev. Lett. {\bf 77} 
(1996) 3865.

\bibitem{troullier}
N. Troullier and J. L. Martins, Phys. Rev. B {\bf 43} (1991) 1993.

\bibitem{klein}
L. Kleinman and D.M. Bylander, Phys. Rev. Lett. {\bf 48} (1982) 1425.

\bibitem{monk} 
H.J. Monkhorst and J.D. Pack, Phys. Rev. B {\bf 13} (1976) 5188.

\end{thebibliography}
\end{document}